\documentclass[showpacs,amsmath,amssymb,aps]{revtex4} 
\usepackage{graphicx}
\usepackage{amssymb}
\usepackage{dcolumn}
\usepackage{bm}

\begin{document}

 \title{Kelvin mode of a vortex \\  in a nonuniform Bose-Einstein condensate}
\author{Alexander L. Fetter}
\email{fetter@stanford.edu}
\affiliation{Geballe Laboratory for Advanced Materials and Departments of Physics and Applied Physics, \\ Stanford University, Stanford, CA 94305-4045}
\date{\today}

 \begin{abstract}  
 In a uniform fluid, a quantized vortex line with circulation $h/M$  can support long-wavelength helical traveling waves $ \propto e^{i(kz-\omega_k t) }$ with the  well-known Kelvin dispersion relation $\omega_k \approx  (\hbar k^2/2M) \ln(1/|k|\xi)$, where $\xi$ is the vortex-core radius.  This result is extended  to include the effect of a nonuniform harmonic trap potential, using a quantum generalization of the Biot-Savart law that determines the local velocity $\bm V$ of each element of the vortex line.  The normal-mode eigenfunctions form an orthogonal Sturm-Liouville set.  Although  the line's curvature dominates the dynamics, the transverse and axial trapping potential also affect the normal modes of a straight vortex on the symmetry axis of an axisymmetric Thomas-Fermi condensate.   The leading effect of the nonuniform condensate  density is to increase the amplitude along the axis away from the trap center.  Near the ends, however, a boundary layer forms to satisfy the natural Sturm-Liouville boundary conditions.  For a given applied frequency,  the next-order correction renormalizes the local wavenumber $k(z)$ upward near the trap center, and  $k(z)$  then increases still more toward the ends.
  \end{abstract}
 
 
 \maketitle

 \section{Introduction}
 
 Quantized vortex lines constitute one of the most characteristic features of superfluidity.  They were originally predicted  to occur in superfluid  He II~\cite{Onsa49,Feyn55} and have  since been the subject of  many studies~\cite{Fett76,Donn91}.  More recently, quantized vortices  have been predicted and observed in the Bose-Einstein condensates of dilute trapped   cold gases~\cite{Matt99,Madi00,Abo01,Halj01,Fett01}.  Apart from the quantized circulation $\kappa = h/M$, their properties  closely resemble those of classical vortex lines, which have long been known to support circularly polarized  oscillatory normal modes~\cite{Kelv80}.  These Kelvin waves have a characteristic long-wavelength dispersion relation $\omega_k \approx (\kappa k^2/4\pi) \ln(1/k\xi)$, where $k$ is the wavenumber along the vortex axis, $\kappa$ is the circulation and $\xi$ is the vortex-core size.  For a classical vortex  with a hollow core, the general dispersion relation involves the  Bessel functions $K_0(k\xi)$ and $K_1(k\xi)$~\cite{Kelv80,Donn91a}, and the preceding  approximate formula applies only for $k\xi\lesssim 0.2$.   In this Kelvin mode,  each element of the vortex core executes a circular orbit with a sense opposite to the circulating velocity around the vortex line.  In the quantum-mechanical version, these modes therefore have quantized angular momentum with $m = -1$ (in units of $\hbar$) relative to the angular momentum of the vortex itself~\cite{Pita61,Isos99}.
 
 In the last year, Bretin {\it et al.\/}~\cite{Bret03} have reported direct  evidence for such Kelvin modes  in an elongated cigar-shaped $^{87}$Rb  condensate.  By driving the condensate with a near-resonant quadrupole excitation ($m =\pm 2$), they observed preferential decay of the mode with $m = -2$ relative to that for $m = 2$.  The authors  ascribed this difference to the creation of two Kelvin-wave  excitations, each with $m = - 1$ and  with opposite momenta $\pm \hbar k$ determined by conservation of energy.  Theoretical analysis confirms this process and the corresponding decay line shape~\cite{Mizu03}.  In addition, transverse images of the expanded condensate show the periodic structure associated with the Kelvin-mode deformation~\cite{Bret03}.
 
Essentially all  theoretical analyses of Kelvin waves in a superfluid have assumed that the fluid is uniform along the vortex axis (see, however, Ref.~\cite{Chev03}) leading to the long-wavelength expression
\begin{equation}\label{kelvin}
\omega_k \approx \frac{\hbar k^2}{2M} \,\log\left(\frac{1}{|k|\xi}\right),
\end{equation}
where the absolute value allows for the possibility of waves propagating in both directions.  This unusual dispersion relation differs significantly from the nondispersive result $\omega_k = ck$ for waves on a string.  It  arises because the dynamics here involves the velocity instead of the more familiar acceleration; in effect, a vortex is massless, and its motion obeys first-order dynamical equations~\cite{Fett67a,Stau68}.  

In the case of a large dilute trapped Bose-Einstein condensate (BEC) in the Thomas-Fermi (TF) limit,  the quantum of circulation $h/M$ and the vortex core size $\xi = (8\pi a n_0)^{-1/2}$ are the only material parameters, where $a$ is the $s$-wave scattering length and $n_0$ is the  central number density of the condensate.  The simplest application is to a  straight vortex initially at rest along the symmetry axis of a condensate.  In this situation,    it would seem natural to replace $n_0$ by  the $z$-dependent  TF density $n(z) = n_0\left(1-z^2/R_z^2\right)$ along the symmetry axis, where $R_z $ is the TF condensate dimension in the $z$ direction. Such a substitution implies that the leading effect of the spatial inhomogeneity appears only in the logarithmic factor in Eq.~(\ref{kelvin}).  One aim of the present work is to demonstrate that there are  other more significant alterations.

A very  general approach would be to rely on the time-dependent Gross-Pitaevskii (GP) equation~\cite{Gros61,Pita61} that determines the time evolution of the condensate wave function $\Psi$.  Such a direct solution for a vortex wave in a nonuniform condensate would involve considerable numerical analysis.  Another possibility is to generalize Pitaevskii's linearized theory of small-amplitude oscillations of a quantized vortex line~\cite{Pita61} to the case of  an inhomogeneous condensate.  This calculation rapidly becomes intricate, for it involves several very different length scales: $\xi$, $ k^{-1}$, $R_\perp$ and $R_z$; in contrast, Pitaevskii studied an unbounded condensate and required only the two parameters $\xi$ and $k^{-1}$.  

A more promising procedure is to rely on the quantum analog of the Biot-Savart law that determines the local velocity of each element of the vortex~\cite{Svid00}.  This formula is readily specialized to the case of a nearly straight vortex line in a large TF condensate.  The system can execute circularly polarized helical waves, and the associated normal-mode eigenfunctions constitute a Sturm-Liouville set subject to natural boundary conditions at the two ends.  For most of the length, the spatial nonuniformity can be incorporated with a phase-integral (WKB) approximation, but a thin boundary layer forms near the ends.    Section II provides the basic formalism, and the approximate solution is presented in Sec.~III.  Comparison with the experimental results and the limitations of the present analysis are discussed in Sec.~IV.

\section{Basic formalism}

Consider an axisymmetric trap potential 
\begin{equation}
V_{\rm tr}(r,z) = \frac{1}{2} M\left( \omega_\perp^2 r^2 + \omega_z^2 z^2 \right),
\end{equation}
where $r$, $\phi$ and $z$ are cylindrical polar coordinates.
 In the TF limit of a large axisymmetric condensate, the particle density is given by 
 \begin{equation}
 n(r,z) = n_0\left( 1 - \frac{r^2 }{R_\perp^2} - \frac{z^2}{R_z^2} \right),
 \end{equation}
 where $R_\perp= \sqrt{2\mu/M\omega_\perp^2}$ and $R_z = \sqrt{2\mu/M\omega_z^2}$ are the radial and axial condensate dimensions, $\mu$ is the chemical potential and $n_0 = M\mu/(4\pi \hbar^2 a)$ is the central density.    A detailed analysis of the time-dependent GP equation for a vortex in such a condensate yields the following expression~\cite{Svid00}  for the velocity ${\bm V}({\bm r},z)$ of an element of the vortex line located at (${\bm r}, z$):
 \begin{equation}\label{vel}
 {\bm V}({\bm r},z) = \frac{\hbar}{2M}\left[ - \ln\left(\xi\sqrt{\frac{1}{R_\perp^2} + \frac{{\cal K}^2}{8}}   \right)\right] \left[ \frac{{\hat {\bm  t}}\times {\bm \nabla} V_{\rm tr}({\bm r}, z)}{\mu \,n({\bm r}, z)/n_0} +{\cal K} {\hat {\bm b}} \right] 
 \end{equation}
 Here, ${\hat {\bm t}}$ is the local tangent vector to the vortex axis,  ${\hat {\bm b}}$ is the binormal and  $\cal K$ is the curvature.  For simplicity, the condensate is assumed to be stationary (nonrotating).
 
 In fact, the argument of the logarithm here requires some discussion.  When Eq.~(\ref{vel}) is used to study the precession of a long straight vortex parallel to the symmetry axis  of a disk-shaped  condensate~\cite{Svid00a}, the curvature $\cal K$ vanishes.  Thus the logarithmic factor reduces to $\log(R_\perp/\xi)$;  it arises  because the long-range circulating velocity field of the vortex line extends to infinity like $1/r$  and is cut off only by the radius $R_\perp$ of the condensate.   
 
 The situation is more complicated when the vortex undergoes a helical distortion like the Kelvin mode.  The original derivation~\cite{Kelv80} of Eq.~(\ref{kelvin}) assumed an arbitrary initial axisymmetric azimuthal velocity  $U_0(r){\hat{\bm \phi }}$  for the unperturbed fluid.  Fortunately,  the situation becomes much simpler for the special case of a straight vortex line with a hollow core of radius $\xi$ in a uniform fluid.  The unperturbed flow velocity   $U_0(r) = \kappa/(2\pi r)$ is  irrotational, and the resulting perturbed motion for the normal  modes can now be characterized by a single scalar velocity potential $\Phi(r,\phi, z, t)$ instead of the three separate components of the perturbations in the velocity.   Indeed, the velocity potential for the Kelvin mode  has the form 
 \begin{equation}
 \Phi(r,\phi, z, t) = \Phi_0\,\exp\left[i\left( kz -\omega t -\phi  \right)\right] K_1(kr),
 \end{equation}
 with angular momentum $m = -1$.   Here $K_1(kr)$ is a Bessel function that decays exponentially at infinity, as is appropriate for a solution of  Laplace's equation that is periodic with wavenumber $k$ along the $z$ axis.  In this picture~\cite{Fett68}, the Kelvin mode 
is simply a surface wave propagating along the vortex  core; the corresponding amplitude is confined to a cylindrical region with radius   $\sim k^{-1}$.  Consequently, for a propagating wave, the natural cutoff inside the logarithm in Eq.~(\ref{vel})  is the smaller of the two lengths $R_\perp$ and $k^{-1}$; thus  a useful approximation for a helical deformation is to replace the logarithm in  Eq.~(\ref{vel}) by 
 \begin{equation}\label{log}
{\cal L}\equiv \ln\left( \frac{R_\perp}{\xi}\,\frac{1}{\sqrt{1+ k^2R_\perp^2}}\right),
\end{equation}
which holds with logarithmic accuracy (note that the curvature $\cal K$ for a small-amplitude helical deformation is proportional to the radius of the deformation and to the squared wavenumber~\cite{Chis}).  

As additional justification for this approximation, it is helpful to recall  the collective analog of Kelvin waves in a rotating vortex lattice with mean areal density $n_v = 2\Omega/\kappa = M\Omega/\pi\hbar$ and intervortex separation $b \equiv (\pi n_v)^{-1/2}$.  The relevant normal modes of this system are characterized by an axial wave number $k$ and have  a dispersion relation~\cite{Gopa64,Fett67,Donn91b}
\begin{equation}\label{coll}
\omega_k \approx \Omega + \frac{\hbar k^2}{2M} \ln\left(\frac{1}{k\xi}\right) + \frac{n_v\kappa}{2} kb\,K_1(kb)\approx
\begin{cases}
(\hbar k^2/2M)\ln\left(1/k\xi \right) ,   &\text{if $kb \gg 1$;} \\
\noalign{\vspace{0.1cm}}
2\Omega+ (\hbar k^2/2M) \ln\left(b/\xi \right),  & \text{if $kb \ll 1$.}
\end{cases}                      
\end{equation}
When $kb$ is large, the wave amplitude arising from a given vortex  vanishes exponentially at the position of the nearest neighbors.  Thus the first  and last terms in Eq.~(\ref{coll}) are negligible, and the dispersion relation reduces to that of a single vortex.  In the opposite limit ($kb\ll 1$), the leading (first)  term  $\approx 2\Omega$ is just the dispersion relation of a classical inertial wave in a uniformly rotating fluid~\cite{Chan61}, and the last term in Eq.~(\ref{coll})  acts to alter the logarithmic factor.  Once again the maximum length in the logarithm is the smaller of the relevant lengths  (which here are $b$ and $k^{-1}$).

\subsection{Dynamical equations for a deformed vortex line}

The preceding equations can now be applied to an originally straight vortex along the axis of symmetry.  The lateral displacement is specified by the two-dimensional vector $x(z,t)$, $y(z,t)$, and only terms of first order in $x$ or  $y$ are retained.  To this order, the tangent vector has the form 
\begin{equation}
\hat {\bm t }\approx \hat {\bm z }+ x' \hat {\bm x} + y'\hat {\bm y },
\end{equation}
where the prime denotes a derivative with respect to $z$.  Similarly, the gradient of the trap potential becomes 
\begin{equation}
{\bm \nabla}V_{\rm tr} = 2\mu\left(\frac{x}{R_\perp^2 }\hat {\bm x } + \frac{y}{R_\perp^2}\hat {\bm y } + \frac{z}{R_z^2}\hat {\bm z} \right).
\end{equation}
Finally, the curvature term becomes ${\cal K}\hat {\bm b } \approx -y''\hat {\bm x} + x''\hat {\bm y }$.  A combination of these various terms with ${\bm V}(z,t) = \dot x \,\hat {\bm x }+ \dot y \,\hat {\bm y }$ yields 
\begin{eqnarray}
 \dot x = \frac{\hbar {\cal L} }{2M}\left[  \frac{2}{1-z^2/R_z^2}\left( \frac{zy'}{R_z^2} -\frac{y}{R_\perp^2} \right) -y''\right] ,\\
  \dot y = -\frac{\hbar {\cal L} }{2M} \left[  \frac{2}{1-z^2/R_z^2}\left( \frac{zx'}{R_z^2} -\frac{x}{R_\perp^2} \right) -x''\right],
\end{eqnarray}
where $\cal L$ is defined in Eq.~(\ref{log}).  

Precessing modes have the form 
\begin{eqnarray}
x(z,t) = f(z)\sin \omega t, \\
y(z,t) = f(z) \cos \omega t.
\end{eqnarray}
Each element of the vortex core executes a clockwise circle (in the negative sense) with angular frequency $\omega$ (assuming $\omega > 0$); for $\omega < 0$, the sense of rotation reverses and becomes counterclockwise (in the positive sense).  The unknown amplitude $f(z)$ obeys the ordinary differential equation 
\begin{equation}\label{geneq}
\omega f = \frac{\hbar {\cal L} }{2M} \left[  \frac{2}{1-z^2/R_z^2}\left( \frac{zf'}{R_z^2} -\frac{f}{R_\perp^2} \right) -f''\right] .
\end{equation}
Here, the  term proportional to $f'$ arises from the tipping of the vortex line and the trap confinement in the axial direction.  The  term proportional to $f$ arises from the trap confinement in the radial direction;  it occurs even for a straight vortex and accounts for the uniform precession of a straight vortex line in a trap.  The  term with $f''$ arises from the  curvature of the vortex line. 

 In the limit of a purely two-dimensional radial trap with $\omega_z \to 0$ and $R_z \to \infty$, the solution is  a uniform propagating wave $f\propto e^{ikz}$ with the dispersion relation
\begin{equation}
\omega_k =\frac{\hbar\,k^2}{2M}\,\ln\left( \frac{R_\perp}{\xi}\,\frac{1}{\sqrt{1+ k^2R_\perp^2}}\right)\left(1-\frac{2}{k^2R_\perp^2}  \right).
\end{equation}
This expression  is {\sl positive\/} for $k^2R_\perp^2  > 2$, reproducing the Kelvin result (\ref{kelvin}) in the limit $kR_\perp \gg 1$ because $k^{-1}$ then provides the relevant logarithmic cutoff.  If, on the other hand, $kR_\perp $ is small, then the frequency becomes {\sl negative\/} and motion approaches a uniform precession in the  positive sense with angular frequency $(\hbar/MR_\perp^2) \ln(R_\perp/\xi)$~\cite{Svid00}, as seen in recent experiments~\cite{Ande00}.  Note that this analysis predicts a stationary helical deformation when the curvature just balances the tendency to uniform precession (namely when $k^2R_\perp^2 = 2$).

\subsection{Sturm-Liouville analysis of normal modes}

In the case of a nonuniform condensate with slowly varying density along the axis, the most interesting normal modes are those with many nodes, describable with a slowly varying wave number.  In this case, the logarithmic parameter $\cal L$ in Eqs.~(\ref{log}) and (\ref{geneq}) is approximately constant.  It is convenient to introduce a dimensionless axial variable $z$  with $R_z$ as length scale, so that $-1 \le z \le 1$.  

The ordinary differential equation (\ref{geneq}) assumes the dimensionless  Sturm-Liouville form 
\begin{equation}\label{SL}
{\cal D} f(z) -2\frac{R_z^2}{R_\perp^2} f(z) = \Gamma^2 \left(1-z^2\right) f(z).
\end{equation}
Here
 \begin{equation}
{\cal D} = -\frac{d}{dz} \left(1-z^2\right) \frac{d}{dz}
 \end{equation}
 is a self-adjoint differential operator and 
 \begin{equation}
\Gamma^2 = \frac{2M R_z^2\omega}{\hbar \ln(1/k\xi)}
\end{equation}
 is a large parameter.  Note that $\Gamma$ is approximately $kR_z$, with $k$ determined from the Kelvin dispersion relation for a uniform fluid (the experimental value $k \approx 0.86 \ \mu{\rm m}^{-1}$  evaluated with $\omega$ as half the quadrupole oscillation frequency for $m = -2$  gives $\Gamma \approx 32$).
 
This equation closely resembles Legendre's equation~\cite{Cour53,Lebe72,Fett80}, and the physical solutions must obey similar ``natural'' boundary conditions at $z = \pm 1$ because of the factor $(1-z^2)$ in the operator $\cal D$.  These conditions mean that $f(z)$ must merely remain finite as $z \to \pm1$.  A series expansion around $z = 0$ leads to a three-term recursion relation, which makes such an approach quite intricate. In addition, the Frobenius method shows that $z = \pm 1$ are regular singular points, with one finite  solution in each case, as anticipated from the natural boundary conditions.  The second solution near $|z|= 1$ has a logarithmic singularity with leading term $\sim \ln(1-|z|)$. The simultaneous pair of finite-value conditions at both ends selects a discrete set of allowed values for $\Gamma^2$, which  is effectively the eigenvalue.

Let $j$ be the discrete index for the allowed solutions.  The real eigenfunctions $f_j(z)$ obey the eigenvalue problem 
\begin{equation}\label{eft}
{\cal D} f_j(z) -2\frac{R_z^2}{R_\perp^2} f_j(z) = \Gamma_j^2 \left(1-z^2\right) f_j(z)
\end{equation}
on the interval $-1 \le z \le 1$, with the eigenvalue $\Gamma_j^2$.  The solutions are orthogonal with 
\begin{equation}\label{ortho}
\int_{-1}^1 dz\,f_i(z) \left( 1-z^2\right) f_j(z) = 0 \quad\hbox{if $i \neq j$}.
\end{equation}

\section{Construction of approximate eigenfunctions and eigenvalues}

To construct the approximate solutions of Eq.~(\ref{eft}), it is necessary to approach the problem in two separate but overlapping regions.  Away from the ends, the WKB method gives a simple description, but the resulting ``outer'' solutions do not satisfy the natural boundary condition.  Instead, the exact solution develops a thin boundary layer (the ``inner'' solution) that requires a different method. Matching the two solutions in the region of common applicability provides the required eigenvalue condition~\cite{Bend78}.

\subsection{Phase-integral (WKB) approximation}

The basic problem is to find an approximate  solution of Eq.~(\ref{eft}) that explicitly includes the density variation that leads to   the factors $1-z^2$.  For an off-center straight vortex in a  disk-shaped condensate, the radial analog of this factor leads to a precession frequency that increases  with increasing  vortex distance  from the axis of symmetry~\cite{Lund00,Fett01}.  Here, in contrast, the density varies {\it along} the direction of the helical wave.  The essential idea is that the properties of the medium vary only on the overall length scale $R_z$. Except for the lowest normal modes, one can therefore define a local axial wavenumber $k$, with  $kR_z\gg 1$.  Based on the experiment of Bretin {\it et al.}~\cite{Bret03}, it is convenient to impose the additional restriction $kR_\perp \gg 1$ (the experimental value is $kR_\perp\approx 4$), and to assume that $k\xi $ is small [the experimental value is $k\xi \approx \frac{1}{4}$, so that $\ln (1/k\xi)\approx 1.4$ is not especially  large].

The phase-integral (WKB) method provides a natural approach to this problem~\cite{Head62,Bend78,Csor98}.  This method assumes that 
\begin{equation}
f(z) = A(z) \exp[i\Gamma S(z)].
\end{equation}
Similar to  the familiar quantum-mechanical case~\cite{Grif95}, the real and imaginary parts of Eq.~(\ref{eft}) yield
\begin{eqnarray}
 \Gamma^2A &=& \frac{2z}{1-z^2} \,A' -\frac{2}{1-z^2}\,\frac{R_z^2}{R_\perp^2}\,A + \Gamma^2 S'^2A - A'' ,\label{re}\\
0 & = & \frac{2z}{1-z^2}\,S'A-2S'\,A' - S''A\label{im}
.  
\end{eqnarray}
The dominant  term arises from the phase $S$, and it is useful to assume the following expansion in inverse powers of $\Gamma$:
\begin{eqnarray}
S &= & S_0 + \Gamma^{-2}S_1 + \cdots,\\
A &= &A_0 + \Gamma^{-2} A_1 + \cdots.\end{eqnarray}
To leading order in $\Gamma$, the real part (\ref{re}) gives \begin{equation}
S_0'^2 = 1,
\end{equation}
so that 
\begin{equation}\label{S0}
S_0(z) = \pm z.
\end{equation}
In this order, the phase is simply that of a  Kelvin wave [in dimensional units, $S_0 = kz$, where 
$k$ is determined as the solution of Eq.~(\ref{kelvin})].

The next-order contribution is from the imaginary part.  Since $S_0''=0$,   Eq.~(\ref{im}) immediately yields 
\begin{equation}
\frac{z}{1-z^2}A_0= A_0',
\end{equation}
which reflects the balance between the longitudinal trap confinement on the left and the curvature on the right.  Direct integration gives 
\begin{equation}
A_0(z) = \frac{1}{\sqrt{1-z^2}}.
\end{equation}
To this approximation, the solution resembles the usual WKB solution~\cite{Grif95}
\begin{equation}\label{lead}
u_0(z) \approx  \frac{1}{\sqrt{1-z^2}}
\begin{cases}
\cos(\Gamma z), &\text{for even solutions;}\\ \noalign{\vspace{0.1cm}}
\sin(\Gamma z), &\text{for odd solutions.}
\end {cases}
\end{equation}
The leading effect of the axial variation caused by the nonuniform trap potential is to increase the amplitude of the normal mode  as $|z|$ increases away from the origin, while  the phase increases uniformly with $z$.

\subsection{Structure of boundary layer near $z = \pm 1$}  

The outer (WKB) solutions (\ref{lead}) apparently diverge for $|z| \to 1$, which violates the natural boundary condition that the amplitude  remain finite.  Thus the exact solution must change character within a thin boundary layer of dimensionless thickness $\delta \ll 1$ to be determined below.  For definiteness, consider the positive end $z = 1$.  It is convenient to introduce an ``inner''  variable $Z$ by the relation
\begin{equation}
Z = \frac{1-z}{\delta},
\end{equation}
which gives an appropriate  boundary layer  near the surface at $z = 1$.

Rewrite the eigenfunction $f(z)$ as $F(Z)$ in terms of the new variable, using $z = 1-Z\delta$, and substitute into Eq.~(\ref{SL}).  This process yields an ordinary differential equation
\begin{equation}\label{temp}
\left(2Z-Z^2\delta\right)\frac{d^2F}{dZ^2} +2\left(1-Z\delta\right)\frac{dF}{dZ} + 2\frac{R_z^2\delta }{R_\perp^2} F + \Gamma^2 \delta^2 \left(2Z-Z^2\delta\right) F = 0.
\end{equation}
Assume an expansion 
\begin{equation}
F(Z) \approx F_0(Z) + \delta F_1(Z) + \cdots,
\end{equation}
and substitute into Eq.~(\ref{temp}).  For small $\delta$, the leading contribution is 
\begin{equation}\label{bessel}
\frac{d^2F_0}{dZ^2} + \frac{1}{Z}\frac{dF_0}{dZ} +\Gamma^2 \delta^2 F_0 = 0.
\end{equation}
It is convenient to make a specific choice of scale factor and take 
\begin{equation}\label{delta}
\delta = \frac{1}{\Gamma}\ll 1
\end{equation}
as the dimensionless boundary-layer thickness.  In this case, Eq.~(\ref{bessel}) reduces to Bessel's equation of order zero, with linearly independent solutions $J_0(Z)$ and $Y_0(Z)$.  The latter is logarithmically singular as $Z\to 0$ and must be rejected, leaving the inner solution that obeys the natural boundary condition
\begin{equation}\label{blsoln}
F_0(Z) = C_0J_0(Z),
\end{equation}
where $C_0$ is a constant.
 
Away from the surface ($Z \gg 1$), the boundary-layer solution (\ref{blsoln}) has the asymptotic form~\cite{Lebea}
\begin{equation}
F_0(Z) \sim C_0\sqrt{\frac{2}{\pi Z}}\,\cos\left( Z - \frac{\pi}{4}\right).
\end{equation}
For comparison, the WKB solution (\ref{lead}) can also be expressed in terms of the inner variable $Z$, using $\sqrt{1-z^2} \approx \sqrt{2Z\delta}$ for $\delta \to 0$.  For the even solution, this procedure yields 
\begin{equation}
f_0(z) \approx \frac{\cos(\Gamma - Z)}{\sqrt{2Z\delta}}.
\end{equation}
Comparison of the overall factors gives the coefficient 
\begin{equation}
C_0 = \pm\frac{\sqrt{\pi\Gamma }}{2}
\end{equation}
using $\Gamma\delta = 1$.  Matching the phase of the two oscillatory solutions fixes the eigenvalue
\begin{equation}
\Gamma_j \approx \left(j+\frac{1}{4}\right) \pi \quad\hbox{for the even solutions},
\end{equation}
where $C_0$ is positive (negative) for  $j$ even (odd).  Similarly, the eigenvalue for the odd solutions is
\begin{equation}
\Gamma_j \approx \left(j+\frac{3}{4}\right) \pi \quad\hbox{for the odd solutions}.
\end{equation}

Figure 1 shows the even and odd solutions for $j = 10$.  The solid lines are the WKB approximations and the dashed lines are the boundary-layer solutions.  Apart from the small discrepancy near the local maxima and minima, the fit is excellent.  In practice, it suffices to take the inner solution $F_0(Z)$  for the interval $0\le Z\le \alpha_1$ where $\alpha_1 \approx 2.4048$ is the first zero of $J_0(Z)$.  For the remaining interval $0\le z \le 1-\delta\alpha_1$, the leading WKB solution $f_0(z)$ gives a good description.

  \begin{figure} 
  \includegraphics[width=3in]{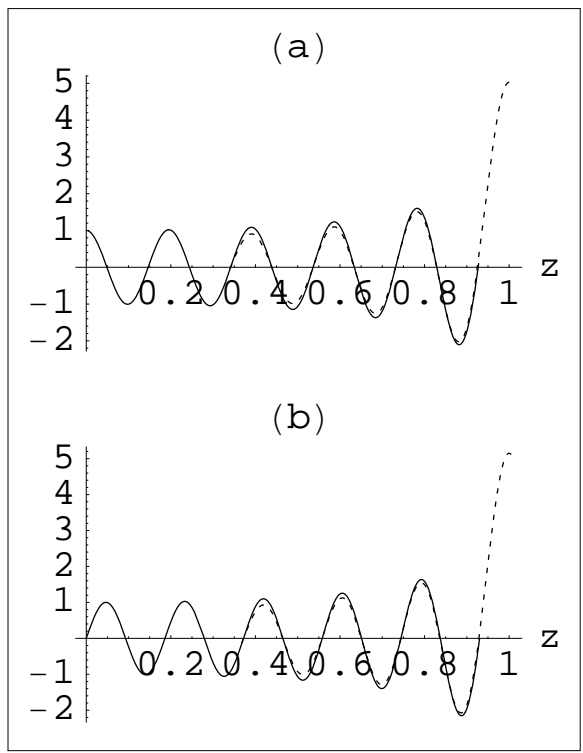}
\vspace{.2in}

 \caption{Typical matched solution for the eigenfunctions corresponding to large excitation quantum number (here for $j = 10$). The solid line is the outer WKB  solution valid away from the ends, and the dashed line is the inner boundary-layer solution.  Curve (a) is the even solution with $\Gamma \approx 10.25\pi$ and curve (b) is the corresponding odd solution with $\Gamma \approx10.75\pi$ (both shown only for $z\ge 0$).}
 \label{fig1}
 \end{figure}

  The normalization integral 
  \begin{equation}
I = \int_{-1}^1 dz\,f_j(z)^2 \left(1-z^2\right) = 2 \int_0^1 dz\,f_j(z)^2 \left(1-z^2\right) 
\end{equation}
  can be evaluated using the same approximations.  Write $I = I_1+I_2$, where
   \begin{equation}
I _1= 2 \int_0^{z_0} dz\,f_j(z)^2 \left(1-z^2\right) \quad \hbox{and}\quad I_2 =  2 \int_{z_0}^1 dz\,f_j(z)^2 \left(1-z^2\right),
   \end{equation}
   with  $z_0 = 1-\delta\alpha_1$.

For $I_1$, use of the WKB approximation immediately gives $I_1\approx 1$ as $\delta\to 0$.  For $I_2$, the boundary-layer solution $F_0(Z) $ readily leads to~\cite{Lebea} 
   \begin{equation}
I_2 \approx  2\delta  \int_0^{\alpha_1} dZ\,\left[ C_0 J_0(Z)\right]^2  \left(2Z\delta\right),
\end{equation}
 which vanishes in the limit $\delta \to 0$ (note that $C_0^2\propto \delta^{-1}$).  
   
   The experiment~\cite{Bret03} effectively drives the vortex with a quadrupole perturbation of given frequency.  Depending on the width of the various resonances, this driving force will excite one or more of the normal modes, and the experiment seems to observe a nearly pure mode.  The appropriate $j$ value would be determined from conservation of energy and the symmetry of the normal modes.  Unfortunately, a more direct observational study of the form of the eigenfunctions  seems impractical.  

\subsection{Next-order WKB description}

It is interesting to consider the next-order contribution for large $\Gamma$, which arises 
 from the first correction to the real part (\ref{re}):
\begin{equation}
2S_0'S_1' = \frac{A_0''}{A_0} -\frac{2z}{1-z^2}\frac{A_0'}{A_0} + \frac{R_z^2}{R_\perp^2}\frac{2}{1-z^2},
\end{equation}
where the last term reflects both the transverse and the axial trap  confinement.  For a wave with $S_0'= 1$, the explicit form of $A_0$ leads to 
\begin{equation}
S_1'(z) = \frac{R_z^2}{R_\perp^2}\frac{1}{1-z^2} +\frac{1}{2(1-z^2)^2},
\end{equation}
and both terms act to increase the phase because $S_1'$ is positive.  Direct integration gives 
\begin{equation}
S_1(z) = \frac{1}{2}\left[\left(\frac{R_z^2}{R_\perp^2} + \frac{1}{4}\right)\ln\left(\frac{1+z}{1-z} \right)+ \frac{1}{2}\frac{z}{1-z^2}\right].
\end{equation}
A combination with Eq.~(\ref{lead}) now includes the effect of the nonuniformity on both the phase and the amplitude
\begin{equation}
f(z) \approx  \frac{1}{\sqrt{1-z^2}}\left\{\begin{matrix}
\cos \\
\sin \end{matrix}\right\}\left( \Gamma z+\frac{1}{2\Gamma}\left[\left(\frac{R_z^2}{R_\perp^2} + \frac{1}{4}\right)\ln\left(\frac{1+z}{1-z} \right)+ \frac{1}{2}\frac{z}{1-z^2}\right]\right).
\end{equation}
An expansion of the phase for small $|z|$ immediately yields the effective dimensionless wave number
\begin{equation}\label{wave}
\Gamma_{\rm eff} \approx \Gamma + \frac{1}{\Gamma} \left(\frac{R_z^2}{R_\perp^2}+\frac{1}{2}\right) > \Gamma,
\end{equation}
where the first term in the correction arises from the overall trap confinement and the second term arises from the combined effect of the curvature and the longitudinal confinement.  Note that $\Gamma_{\rm eff}$ exceeds $\Gamma$ obtained as the solution of Eq.~(\ref{kelvin}).

More generally, the quantity $S'(z)$ defines the effective dimensional wavenumber $k_{\rm eff} (z)$ through the relation
\begin{equation}
S'(z) =  k_{\rm eff} (z)R_z   \approx kR_z\left[ 1 + \frac{1}{k^2R_\perp^2 }\frac{1}{1-z^2} + \frac{1}{2k^2R_z^2 }\frac{1}{(1-z^2)^2}\right],
\end{equation}
where $k$ is the solution of Eq.~(\ref{kelvin}) (ignoring the slow variation of the logarithmic factor).  For small $|z|$, this expression reduces to (\ref{wave}), but it now includes the region where $1-z^2$ is not small.  Eventually this approach will fail near the ends of the condensate, but it does indicate that the Kelvin wave in a TF condensate should have not only an increased amplitude but also an increased wavenumber as $|z|$ increases from the center of the trap along the symmetry axis.

\section{Discussion and potential comparison with experiment}
As noted in Ref.~\cite{Bret03}, the Kelvin dispersion relation (\ref{kelvin}) for the ENS experiments predicts a  wavenumber $k\approx 0.86\,\mu{\rm m}^{-1}$.  This value agrees well with the observations after accounting for the expansion of the condensate (which is strong in the radial direction but only weak in the axial direction for the ENS trap, where $R_z/R_\perp\sim 8.3$).  Since $\Gamma \approx 32$,  the corrections studied in the present paper are formally small (of order $\Gamma^{-2}$).  Note, however, the important effect of trap anisotropy through the factor $R_z^2/R_\perp^2 \sim 70$ in Eq.~(\ref{wave}).  As a result, the correction is instead of order $(kR_\perp)^{-2}\sim 0.07$, which might well be observable.  

Unfortunately, any direct comparison with Ref.~\cite{Bret03} is suspect because the original unperturbed vortex  was  bent.  Such behavior was predicted in~\cite{Garc01a,Garc01b,Afta01} and subsequently observed in~\cite{Rose02}.   In contrast, the present theoretical description assumes a straight unperturbed vortex line along the axis of symmetry, and an extension to an initially bent vortex line would be difficult.  

A straight vortex line has been proved to be the ground state for a spherical and a flattened disk-shaped trap with $R_z/R_\perp \le 1$~\cite{Afta02}.  
More detailed numerical work~\cite{Modu03} finds that the bent vortex first becomes the stable state for $R_z/R_\perp\gtrsim 2$, which agrees with the observations in~\cite{Rose02} where $R_z/R_\perp\sim 8.2$.  Thus it would be interesting to study Kelvin waves in a condensate that is only slightly elongated, in which case the present theoretical description should apply.

\acknowledgments
I am grateful to J.~Dalibard, S.~Stringari, and A.~Svidzinsky for stimulating discussions that initiated this work.

 \end{document}